\documentstyle[psfig,prb,aps]{revtex}
\begin{document}
\draft
\twocolumn[\hsize\textwidth\columnwidth\hsize\csname
@twocolumnfalse\endcsname

\title{Errors in Hellmann-Feynman Forces 
due to occupation number broadening, and how they
can be corrected}

\author{F. Wagner, Th. Laloyaux and M. Scheffler}
\address{Fritz-Haber-Institut der Max-Planck-Gesellschaft, 
Faradayweg 4-6,
D-14195 Berlin-Dahlem, Germany}
\date{August 4, 1997}
\maketitle
\begin{abstract}
In {\sl ab initio} calculations of electronic structures, total energies, and 
forces, it is convenient and often even necessary to employ a broadening
of the occupation numbers. If done carefully, 
this improves the accuracy of the calculated
electron densities and total energies and stabilizes the convergence
of the iterative approach towards self-consistency.
However, such a boardening may lead to an error in the calculation of the
forces.
%, and it is unclear, how significant this error will be and how it
%should be corrected.
%, but in standard schemes 
%it results in incorrect forces.
Accurate forces are needed for an efficient 
geometry optimization of polyatomic systems and for {\em ab initio} 
molecular dynamics (MD) calculations.
The relevance of this error and possible
ways to correct it will be discussed in this
paper.
The first approach is computationally very simple and in fact exact
for small MD time steps.
This is demonstrated  for the example of the
vibration of a carbon dimer and for the relaxation 
of the top layer of the (111)-surfaces of aluminium and platinum.
The second, more general, scheme employs 
linear-response theory and is applied to the calculation 
of the surface relaxation of Al\,(111).
We will show that the quadratic dependence
of the forces on the broadening width enables an efficient
extrapolation to the correct result.
Finally the results of these correction methods will be 
compared to the forces obtained by using the
smearing scheme, which has been proposed by Methfessel and Paxton.
\end{abstract}

\vskip2pc]

% body of paper here
In {\sl ab initio} electronic structure and total energy
calculations  the integrals over the Brillouin zone 
are commonly replaced by the sum over a 
mesh of  {\bf k}-points\cite{Chadi,Monkhorst}.
This approach is very efficient for insulators, 
but for metallic systems
convergence with respect to the number of {\bf k}-points becomes slow.
Here the introduction of fractional occupation numbers is a 
convenient way to improve the {\bf k}-space integration
and in addition to stabilize the convergence
in the iterative approach to self-consistency.
In these broadening schemes the eigenstates are occupied according to
a smooth function, e.g.\ a Gaussian \cite{FuHo} or the Fermi 
function \cite{Mermin,Davenport,gillan,jn} at a finite temperature.

When a broadening scheme is employed in a density 
functional theory calculation,
the computed electron density of the ground state $n_0({\bf r})$ does not
minimize the functional of the total energy $E$ but 
the functional of the free energy $A$:

\begin{equation}
A\left[T_{\text{el}};n \right] = E\left[T_{\text{el}};n \right]
 - T_{\text{el}} S\left[T_{\text{el}};n \right]\quad,
 \label{emts}
\end{equation}
where $S$ denotes the entropy associated with 
the occupation numbers of the Kohn-Sham orbitals
and $T_{\rm el}$ is the broadening parameter.
In the case of Fermi broadening\cite{ashcroft} we get:
\begin{equation}
S =- k_{B} \sum_{i} \left[f_i \ln f_i + \left(1-f_i\right)
\ln \left(1-f_i\right)\right]\quad.
\label{entropy}
\end{equation}

Since the temperatures commonly used for the broadening
are much higher than the
physical ones (it is convinient to use
$k_{B}T_{\text{el}}\sim 0.1~\text{eV}$),
neither the total energies nor the free energies (Eq.~\ref{emts})
are directly meaningful.

\begin{figure*}[b]
\includegraphics{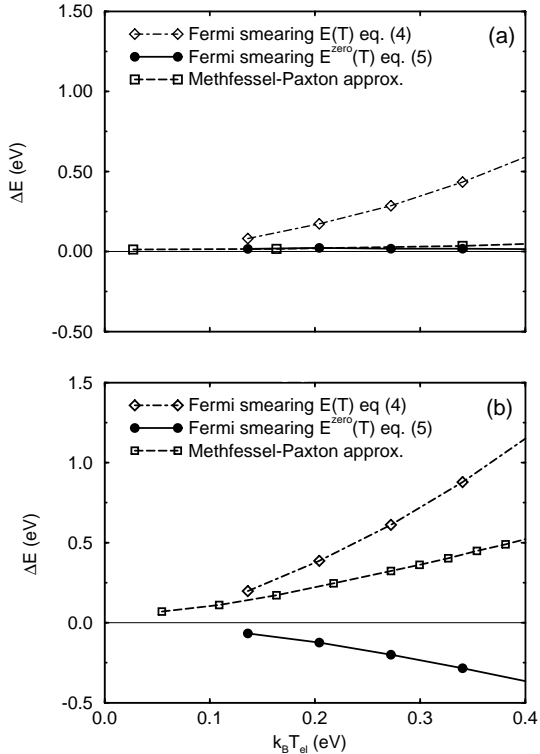}
\caption{
Total energy $E(k_{\rm B}T_{\rm el})$ (open diamonds, dash-dotted line)
and energy extrapolated to $k_{\rm B}T_{\rm el}=0\: {\rm eV}$ $E^{\rm zero}$ 
(filled circles, solid line)
as a function of $k_{\rm B}T$ using a Fermi broadening and
total energy $E^{\rm tot}$ (open squares, dashed line) using the
Methfessel-Paxon smearing of first order for a slab
of four-layers of aluminuim (a) and platinum (b).
Dots present computed values, lines are fits to guide the eye.}
\label{al:pt:ene:temp}
\end{figure*}

One way to obtain the ground state energy at
zero temperature is based on the well known fact\cite{ashcroft}, 
that for the free electron gas, 
the quantities $A$ and $E$ depend quadratically 
on $T_{\text{el}}$.
%Therefore the harmonic aprroximation should be
%valid at least for nearly free electron like systems:
Therefore one can write:
\begin{eqnarray}
A\left(T_{\text{el}}\right)=E^{\text{zero}}
-\frac{1}{2}\gamma^{2}T_{\text{el}}^{2}
+O\left(T_{\text{el}}^{3}\right)\quad
\label{free} \\
E\left(T_{\text{el}}\right)=
E^{\text{zero}}+\frac{1}{2}\gamma^{2}T_{\text{el}}^{2}
+O\left(T_{\text{el}}^{3}\right)\quad.
\label{energy}
\end{eqnarray}
As it was pointed out by Gillan\cite{gillan},
it follows from Eqs.~(\ref{emts}), (\ref{free}) and
(\ref{energy}) that the extrapolation of the total
energy towards the $T_{\text{el}}$=0 result is:
\begin{eqnarray}
E^{\text{zero}}&=&E(T_{\text{el}}\rightarrow0)=
A(T_{\text{el}}\rightarrow0)\\
&\approx& E(T_{\text{el}})-\frac{1}{2}T_{\text{el}}
S(T_{\text{el}})\quad.
\label{ezero}
\end{eqnarray}
Obtaining $E^{\text{zero}}$ using Eqs.~(\ref{ezero}) 
and (\ref{entropy})
is straightforward and gives very satisfactory results\cite{jn}.
%cfw
This is shown in Figure \ref{al:pt:ene:temp}(a) for a slab consisting of 
four layers of Aluminium. Here the extrapolation (filled circles)
matches perfectly the
zero-temperature energy, even for quite large broadening temperatures.
For a systems like Platinum, which was chosen as an example for a not
free electron like system, Figure \ref{al:pt:ene:temp}(b) shows, 
that this extrapolation
is indeed an approximation. But if the broadening parameter is chosen
carefully (typically used broadenig parameters are about 0.1 eV or lower)
the extrapolation still gives acceptable results.

%cfw
Calculating the forces, however, is more complicated.
The forces on atoms are defined as the derivative 
of the total energy $E^{\text{zero}}$ with respect to the atomic positions:
\begin{eqnarray}
{\bf F}^{\text{zero}}& =&
      -\frac{\partial E^{\text{zero}}}{\partial{\bf R}}\\
{\bf F}^{\text{zero}}& =&-\frac{\partial A}{\partial{\bf R}} 
   -\frac{1}{2}T_{\text{el}}
\frac{\partial S}{\partial{\bf R}} 
\label{fzero}
\end{eqnarray}
\vspace{9cm}

\begin{figure}
\includegraphics{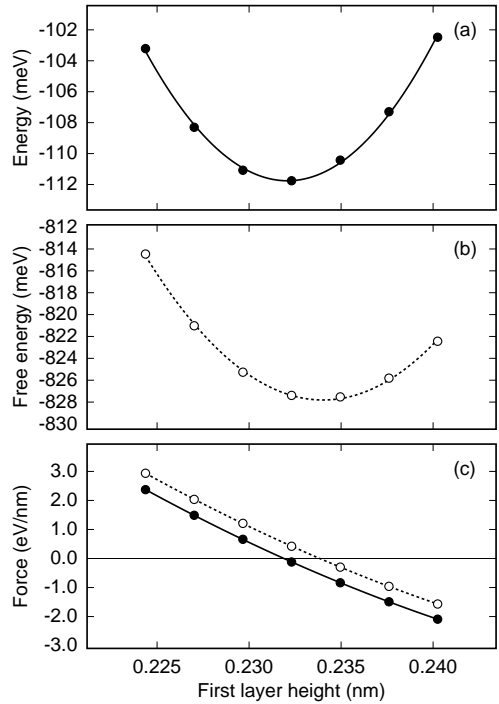}
\caption{(a) Zero-temperature energy $E^{\text{zero}}$ 
and (b) free energy $A$ of a four-layers Al~(111) slab
as a function of the height $Z$ of the first atomic layer.
To demonstrate the effect, a Fermi-Dirac broadening with 
an untypically high broadening pararmeter $k_B T_{\text{el}}$=0.5~eV 
has been used.
(c) Force acting on the first atomic layer
(open dots are non-corrected and full dots are corrected forces).
Dots present computed values, lines are fits to guide the eye.}
\label{relax}
\end{figure}

While the quantity
\begin{equation}
{\bf F}^{\rm HF}(T_{\text{el}})
= -\frac{\partial A}{\partial{\bf R}}
\label{ftel}
\end{equation}
is easily evaluated, the evaluation of
 \( \partial S / \partial{\bf R} 
\) is somewhat more elaborate.
Neglecting the entropy term in 
Eq.~(\ref{fzero}) 
implies that the 
forces are not in agreement with the gradient 
of the total energy of
Eq.~(\ref{ezero}).
As a consequence, when those forces are used 
to relax the atoms towards 
their equilibrium
positions, the obtained geometry is likely to be 
different from that 
which minimizes $E^{\text{zero}}$.
Figure \ref{relax} presents results for a four-layers
aluminum (111) slab, with fully separable, norm-conserving
pseudopotentials\cite{hamann,gss},
a plane-wave basis set ($E^{\text{cut}}=8~\text{Ry}$) and
18 $\bf k$-points\cite{chadi} to sample the surface Brillouin zone.
An untipically high broadening of $0.5~\text{eV}$ was used 
to show the effect and it is clearly visible
that the Hellmann-Feynmann-Force 
$ {\bf F}^{\rm HF}(T_{\text{el}})$
acting on the surface
layer vanishes for the position minimizing the free energy,
as it should according to Eq.~(\ref{ftel}).
But the minimum of the total energy is at a different position.\\
Even at this untypically high broadening temperature,
which was chosen to illustrate the effect, the error
in the equilibrium position is only $10^{-3}$ nm,
which is less than one percent of the interlayer distance.
This indicates, that the error in
the forces due to occupation number\\

\vspace{9cm}

 broadening may be 
neglegible in the case of relaxations if the temperature 
is chosen not unreasonable high, but unfortunally this has to be checked for
each particular system.

%Dynamics
The situation is less clear and the needs are more demanding, when it
comes to molecular dynamics (MD) simulations.
When the non-corrected forces (Eq.~\ref{ftel}) are used  to perform MD,
the sum of potential kinetic
energies of the atoms is not a conserved quantity.
An illustration of this is shown in Figure \ref{md}(a)
for a MD simulation of the vibration of a carbon dimer.
The graph shows the total energy (open circles) and the free energy (closed
circles) of the systems as a function of the interatomic distance.
The trajectory was integrated over approx.\ two periods using the verlet 
algorithm and no thermostat had been employed.
Using the non-corrected forces produces a motion in which the total energy 
is not conserved but oscillates with the frequency of the motion.
Figure \ref{md}(a) cleary shows, that at the turning points,
where the kinetic energy vanishes, the potential energies are
different, which is obviously unphysical.
In fact, the free energies, given by 
Eq.~(\ref{emts}), at the turning points are equal.

\begin{figure}
\vspace{10.5cm}
\includegraphics{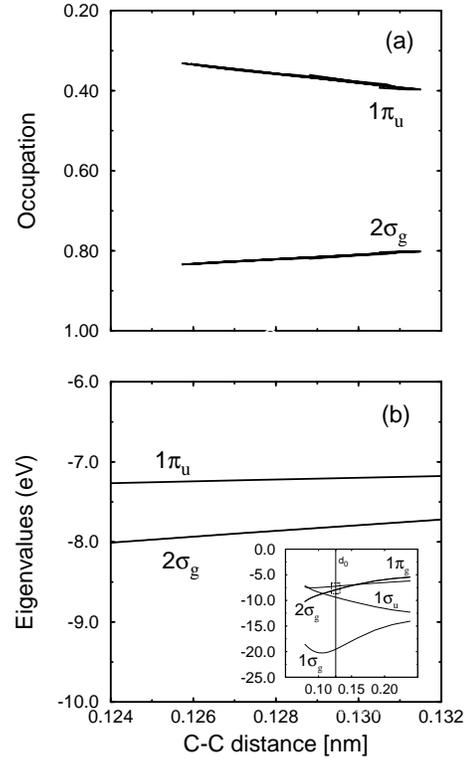}
\caption{
(a) Occupation numbers of the highest occupied molecular orbital 
$1\sigma_{\rm g}$ (HOMO, two-fold degenerate)
and the lowest unoccupied molecular orbital (LUMO) 
$2\pi_{\rm u}$
as a function of the interatomic distance
during the vibration of a carbon dimer, 
using $k_B T_{\text{el}}$=0.5~eV.\\
(b) Energy levels of a carbon dimer as a function
of the interatomic distance. $d_0$ marks the quilibrium 
distance.
 }
\label{occ}
\end{figure}

Plotting the eigenvalues as a function of the
intermolecular distance (Fig.~\ref{occ}(b)) shows
a crossing of the  $2 \sigma_g$ level
and the two-fold degenerated $1 \pi_u$ levels near\\
\vspace{9cm}

\includegraphics{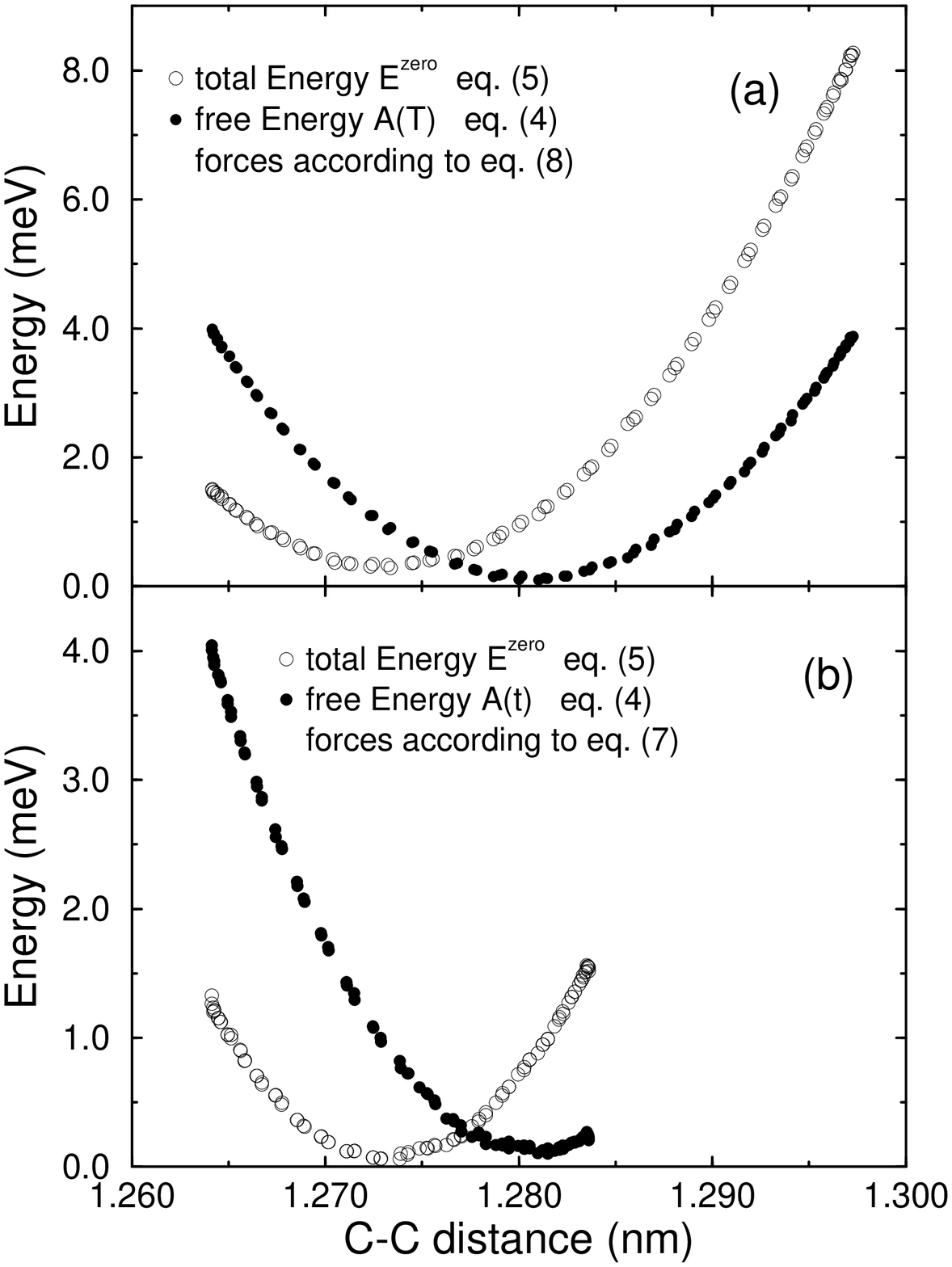}
\begin{figure}
\caption{Free Energy $E^{\text{zero}}$ 
(filled circles) and total energy $E$ (open circles) 
as a function of interatomic distance 
during the  vibration of a carbon dimer.
The points are taken at different times during the vibration.
The border points at the outer left and right are the turning points
of the motion, thus the range of distances covered in each figure correspond 
to the amplitude  of the vibration.
(a) Non-corrected forces have been used  for the molecular dynamics.
Obviously the potential energy at the two turning points is very different.
(b) Corrected forces have been used for the molecular dynamics.
The potential energy at the two turning points is nearly the same.}
\label{md}
\end{figure}

the equilibrium distance $d_0$.
Since in the groundstate, the $2 \sigma_g$ level is empty (LUMO), while the
$1 \pi_u$ levels are fully occupied (HOMO),
the introduction of a broadening function causes a
noticeable change in the occupation numbers
of these orbitals (Fig.~\ref{occ}(a)) during the vibration.
This leads to a non-negligible correction to \( {\bf F}(T_{\text{el}}) \) 
according to Eq.~\ref{fzero}.
We will now describe how this
entropy contribution to the forces 
can be evaluated.
From Eq.~(\ref{entropy}), one obtains the expression for
$T_{\text{el}} \partial S / \partial {\bf R}$:
\begin{equation}
T_{\text{el}}\frac{\partial S}{\partial {\bf R}}=
-2 k_B T_{\text{el}}\sum_{i} 
\frac{\partial f_i}{\partial{\bf R}} \ln
\frac{f_{i}}{1-f_{i}}\quad,
\end{equation}
which, in the case of a Fermi distribution 
for the occupation numbers,
reduces to
\begin{equation}
T_{\text{el}}\frac{\partial S}{\partial {\bf R}}=
2\sum_{i} \frac{\partial f_i}{\partial{\bf R}}\epsilon_{i}\quad.
\label{fdcorr}
\end{equation}

The $\epsilon_{i}$ denote the energies of the Kohn-Sham orbitals.
Using a Fermi distribution for the \( f_{i} \), 
we find, that in the case of the vibrating dimer
on the relevant length scales 
the occupation numbers change linearly with {\bf R}
(see Fig.~\ref{occ}(a)), i.e.\
\( \partial f_{i}/\partial {\bf R}$
is nearly a constant.
Figure \ref{md}(b) illustrates the symmetric 
and total energy-conserving
vibrations obtained when the forces are corrected 
according to Eqs.~(\ref{fzero}) and (\ref{fdcorr})
using the linear treatment, 
which is well satisfied for the small time steps which are
typically used in a MD simulation.
We expect that for many {\em ab initio} MD calculations 
a good estimate
of $\partial f_i / \partial {\bf R}$ can be obtained 
from the past atomic geometries.
This is particularly true because due to the small time steps 
($\Delta t \approx \frac{1}{20}$ of the period of the vibration)
the difference between adjacent 
geometries in a MD calculation is typically very small, 
i.e. only about $10^{-4}$ nm.

%Linear Response
To complete our analysis, we also applied a more general method,
for calculating  $\partial f_i / \partial{\bf R}$.
We will use again the Al slab as an example.
At constant temperature and number of electrons, 
the occupation numbers
$f_i$ depend on the one-electron energies $\epsilon_i$ 
and on the chemical
potential $\mu$.  Thus,
\begin{equation}
\frac{\partial f_i}{\partial{\bf R}}=
\frac{\partial f_i}{\partial \epsilon_i}
\left(\frac{\partial\epsilon_i}{\partial{\bf R}}-
\frac{\partial\mu}{\partial{\bf R}} \right)\quad,
\label{dfidr}
\end{equation}
in which the chemical potential is obtained from the
constraint that the $f_i$ sum up to the number of electrons.
Because a given atomic displacement will result 
in an increase of some 
eigenvalues and a decrease of the other ones, 
the derivative of the 
chemical potential is smaller than those of the eigenvalues.
It is nevertheless not negligible: 
in the case of the relaxation of Al\,(111), we find 
that the contribution
of the derivative of the chemical potential is about 0.2 times
that of the derivative of the eigenvalues. \\
Linear-response theory\cite{baroni1,baroni2,xg} 
enables us to calculate
the quantities $\partial\epsilon_i / \partial{\bf R}$, which
are the expectation values of $\partial \cal H / \partial{\bf R}$
for the eigenstates $\left|\psi_i\right\rangle$
\begin{equation}
\frac{\partial\epsilon_i}{\partial{\bf R}}=
\left\langle\psi_i\left|\frac{\partial \cal H}{\partial{\bf R}}
\right|\psi_i\right\rangle\quad.
\end{equation}
Other methods would be more time-consuming 
and thus inadequate for the purpose
of geometry optimization or MD, where the forces 
must be calculated for
many different atomic configurations and sometimes 
for systems of
100 atoms and more.
The dielectric matrix, for example, would require 
the calculation of many
bands and the inversion of large matrices. \\
The explicit dependence of $\cal H$ on $\bf R$ is only in
$V_{\text{ext}}$, but $V_H$ and $V_{\text{xc}}$ depend on the
atomic positions through the electron density.
For that reason,
an evaluation of the forces not using the values 
of the $\epsilon_i$
of previous atomic positions requires a self-consistent calculation 
of the derivatives
$\partial\epsilon_i / \partial{\bf R}$.
Neglecting the dependence of $V_H$ and $V_{\text{xc}}$ 
on $\bf R$ to 
avoid the use of an iterative computation would result in 
unacceptable errors: in the example of the Al slab, 
we encountered 
cases where non self-consistent forces were too large 
by an order of 
magnitude.  \\
The derivative of the electron density is given by
\begin{equation}
\frac{\partial n({\bf r})}{\partial{\bf R}} =
2\sum_{i} \left\{
\frac{\partial f_i}{\partial{\bf R}} 
\left| \psi_i({\bf r})\right|^2
+2f_i \Re\left[\psi_i^*({\bf r})
\frac{\partial\psi_i({\bf r})}{\partial{\bf R}} \right]
\right\}\quad,
\label{dens}
\end{equation}
where the first term accounts for the redistribution 
of the electrons among
the orbitals due to the variation of the $\epsilon_i$ 
and the second term
comes from the modification of the wavefunctions.
From the current approximation 
of $\partial\epsilon_i / \partial{\bf R}$,
Eq.~(\ref{dfidr}) is used to calculate the first term.
The second term requires the
resolution of
\begin{equation}
\left({\cal H}-\epsilon_i\right)
\left|\frac{\partial\psi_i}{\partial{\bf R}}\right\rangle=
-\left(\frac{\partial {\cal H}}{\partial{\bf R}}-
\frac{\partial\epsilon_i}{\partial{\bf R}}\right)\left|\psi_i
\right\rangle\quad.
\label{dpsi}
\end{equation}
Equation~(\ref{dpsi}) is ill-conditioned, because the operator
$\left({\cal H}-\epsilon_i\right)$ of the left-hand side 
has in general
eigenvalues of either sign, some of them having 
small absolute values.
Later in this paper (Eqs.~\ref{init} and \ref{iter}), 
we are going to explain an iterative resolution method
instrumental for positive-definite operators 
and which converges better
if the eigenvalues of that operator are large.
In order to make that method applicable to our problem, 
we separate 
the Hilbert space in two subspaces: 
the first one is spanned by the computed
eigenstates,
and the second one is its complementary subspace
(spanned by unoccupied states).
In practical calculations, only the $i_{\text{max}}$
lowest energy levels, including
all occupied states and some unoccupied states, 
are computed and the 
occupation numbers are fractional only 
for the levels with energies \( \pm 
k_{B}T_{\text{el}} \) around the Fermi level.
Let $\cal P$ be the projector on the second subspace 
($i>i_{\text{max}}$)
and $\left|\partial\tilde{\psi}_i / \partial{\bf R}\right\rangle$ be
${\cal P} \left|\partial\psi_i / \partial{\bf R}\right\rangle$.
Now, we can rewrite Eq.~(\ref{dens}) as
$$
\frac{\partial n({\bf r})}{\partial{\bf R}}=
2\sum_{i=1}^{i_{\text{max}}} \left\{
\frac{\partial f_i}{\partial{\bf R}} \left| \psi_i({\bf r})\right|^2+
\right.$$
$$
\left.
2\sum_{j=i+1}^{i_{\text{max}}} 
\frac{f_i-f_j}{\epsilon_i-\epsilon_j}
\Re \left[ \left\langle\psi_i\left|
\frac{\partial H}{\partial{\bf R}}
\right|\psi_j\right\rangle
\psi_i({\bf r}) \psi_j({\bf r})^* \right]\right. 
$$
\begin{equation}
\left. +2f_i \Re\left[\psi_i({\bf r})
\frac{\partial\tilde{\psi}_i({\bf r})^*}{\partial{\bf R}} \right]
\right\}\quad.
\label{dens2}
\end{equation}
Some of the differences $(\epsilon_i-\epsilon_j)$ appearing
in the denominator of the second term in Eq.~(\ref{dens2}) are
small but, in that case, 
the difference between the occupation numbers
in the numerator is small as well, so the whole fraction 
has a finite value. \\
The quantities
$\left|\partial\tilde{\psi}_i / \partial{\bf R}\right\rangle$
are the solutions of
\begin{equation}
\left({\cal H} -\epsilon_i \right)
\left|\frac{\partial\tilde{\psi}_i}{\partial{\bf R}}
\right\rangle=
-{\cal P}\frac{\partial\cal H}{\partial{\bf R}}
\left|\psi_i\right\rangle\quad.
\label{noncom}
\end{equation}
Since Eq.~(\ref{noncom}) is only 
for the subspace $i>i_{\text{max}}$,
the operator $\left({\cal H} -\epsilon_i \right)$ 
is positive-definite.
Moreover, an approximation $\cal D$ to
$\left({\cal H} -\epsilon_i \right)$ can be obtained
by neglecting its off-diagonal matrix elements.  
In the plane-wave basis
set we use in the examples, this amounts to including 
the kinetic part of
the hamiltonian and the average value of the effective potential.
The algorithm introduced by Williams and Soler\cite{ws}
 can be generalized 
to solve Eq.~(\ref{noncom}) iteratively.
An initial guess is given by
\begin{equation}
\left|\frac{\partial\tilde{\psi}_i^{(0)}}
{\partial{\bf R}}\right\rangle=
-{\cal D}^{-1} {\cal P}
\frac{\partial\cal H}{\partial{\bf R}}\left|\psi_i\right\rangle
\label{init}
\end{equation}
and the sequence of approximations is obtained using
\begin{eqnarray}
\left|\frac{\partial\tilde{\psi}_i^{(n+1)}}
{\partial{\bf R}}\right\rangle=
\left|\frac{\partial\tilde{\psi}_i^{(n)}}{\partial{\bf R}}
\right\rangle +
{\cal P} {\cal D}^{-1} \left( 1-e^{-\Delta{\cal D}} \right)\\
\left[ (\epsilon_i -{\cal H})
\left|\frac{\partial\tilde{\psi}_i^{(n)}}
{\partial{\bf R}}\right\rangle -
{\cal P}
\frac{\partial\cal H}{\partial{\bf R}}\left|\psi_i
\right\rangle \right]
\quad.
\label{iter}
\end{eqnarray}
In order to make the convergence as fast as possible,
the largest value keeping the algorithm stable is chosen
for the constant $\Delta$.
The operators ${\cal D}^{-1}$ and $e^{-\Delta{\cal D}}$ 
are easily computed,
because ${\cal D}$ is diagonal.
The operator
$\partial {\cal H} / \partial {\bf R}$ is different 
at each iteration,
because the value of
$\partial n({\bf r}) / \partial {\bf R}$ 
on which it depends is updated
each time, in order to make the charge redistribution 
converge to
self-consistence.  This is analogous to the method used
for the total-energy minimization\cite{payne}, 
where the convergence towards
the self-consistent charge density and the diagonalization of the
hamiltonian are performed simultaneously. \\
The method explained above was applied to the Al slab.
The corrected force is plotted in Figure \ref{relax}(c) 
and it indeed
vanishes for the geometry that minimizes the zero-temperature
total energy.

%quadratic Extrapolation
As expected from Eqs.~(\ref{ftel}) and (\ref{free}),
the dependence of the non-corrected force on $T_{\text{el}}$ 
is quadratic.
Therefore it should be possible to extrapolate the
force to the value at zero temperature from two
points at finite temperature.
Figure \ref{al:pt:ene:temp} (a) shows the result for the Al\,(111) slab,
using 48 {\bf k}-points in the irriducible part of the
Brillouin zone.
Using the formula
\begin{equation}
{\bf F}^{\text{zero}} =
\frac{dA(T_1)/d{\bf R}-(T_1/T_2)^2 dA(T_2)/d{\bf R}}
{(T_1/T_2)^2-1}
\label{extrap}
\end{equation}
we obtain a value which matches quite well the 
nearly constant value, obtained 
for the entropy-corrected force and by the linear
treatment.
This extrapolation scheme even works for metals
which are not free electron like, as it is shown
in Figure \ref{al:pt:ene:temp} (b) for the case of platinum.
In the case we used 69 {\bf k}-points in the irriducible
part of the Brillouin zone.
As for a transition metal no broadening temperature
above 0.1-0.2 eV should be used to retain the
properties of the Fermi surface, in many cases 
there might be no need to correct the forces at all.
But calculating the forces at two different
temperatures and extrapolating to $T_{\rm el}=0$
may lead to some additional improvement.

%MethfesselPaxton
Methfessel and Paxton\cite{meth} proposed an improved smearing scheme
in which the orbitals are occupied according to a smooth approximation
of the step function.
The computed quantities (energy, forces,\ldots) 
converge towards their zero-temperature values 
when the order of the approximation is increased.  
Unfortunately higher-order approximations become 
more and more wigglier and therefore require larger 
$\bf k$-point sets.
In practical calculations therefore only the first order approximation is
used. In Figure \ref{al:pt:ene:temp} the differences of the calculated 
energies for finite temperatures to the energy at zero temperature for
the Al-111 (a) and the Pt-111 (b) slab is plotted as a function of the
broadening parameter.
For both systems the deviation of the energies obtained by using the first 
order Methefessel-Paxton scheme (full circles) ist comparable to that of the 
extrapolated energies (open squares) using the finite temperature approach 
(open diamonds).

Figure \ref{al:pt:for:temp} shows that for both systems the error in the 
forces which are obtained when the Methefessel-Paxton scheme is used
also is comparable to the error in the forces obtained by the quadratic
extrapolation.
But Figure \ref{al:pt:for:temp} also shows, that the forces obtained by
using
the MP-scheme are wigglier, especially for the Al-111 slab, where only 
48 {\bf k-}Points have been used. 

\vspace{10cm}

\includegraphics{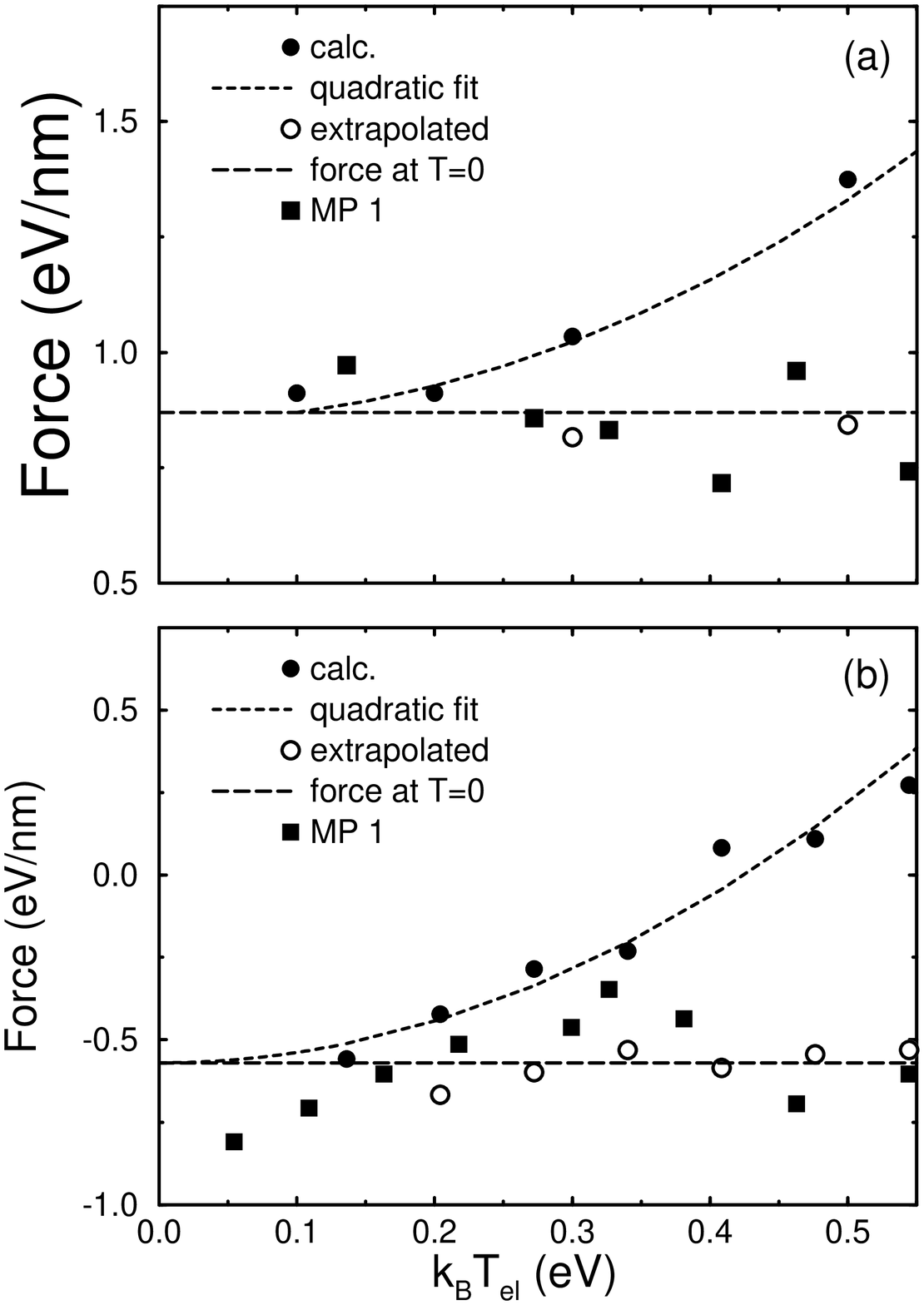}

\begin{figure}
\caption{
Non-corrected force acting on the first layer of a four layer (111) slab 
of aluminium (a) and platinum (b) as a function of
the broadening energy $k_B T_{\text{el}}$
using a fermi-smearing (closed circles) and the Methfessel-Paxton-scheme
of first order (closed squares).
Parabolic fit (dotted line) and forces extrapolated
to T=0 (open markers).
Dashed line corresponds to the minimum of the parabola.
}
\label{al:pt:for:temp}

\end{figure}

%Conclusion
In conclusion, we have shown how the entropy arising 
from the broadening
of the occupation numbers can be included 
in the calculation of the forces.
For small distortions the dependences 
of the occupation numbers are 
linear to a good approximation.
Thus, information from the history of the MD run 
can be used to 
determine \( \partial f_{i}/\partial {\bf R} \).
For the general case, e.g. when the needed information is not 
available from the history, we developed 
an iterative method based on
the linear-response theory.
The forces obtained in this way are the exact derivatives
of the extrapolated zero-temperature energy.
Neglecting the contribution of $V_{H}$ and $V_{\text{xc}}$ to
\( \partial \epsilon_{i}/\partial {\bf R} \), 
in other words stopping 
the calculation after the first iteration would result 
in unacceptably errors.
If $\bf k$-point convergence is fulfilled, 
the corrected force is nearly
independent of the broadening temperature 
in a wide temperature range.
In that case, extrapolation to zero temperature 
from the results at
two finite temperatures also gives good results.  \\
Among the three methods presented in this paper, 
linear-response theory
is the most general one: it is valid 
for any dependence of the
one-electron energies on the atomic positions.
On the other hand, this method is computationally very
expensive, which limits it's usefuleness in
practical calculations.
The extrapolation method
requires the calculation of the forces 
for two different temperatures,
which in general does not double the number of 
iterations, since the number of iterations needed is
much smaller starting from a selfconsistent
charge density at a different temperature.
Therefore extrapolation from two 
different temperatures is more practical, especially 
if the number of degrees of freedom is large.\\
While the error in the forces is small compared
to the usual accurancy in atomic relaxations,
and therefore might be neglected if the broadening
temperature is choosen carefully,
the correction is especially important 
to stay consistent in a molecular dynamics simulation,
where the quantity to be conserved should be
the ground state energy of the electronic
system and not the unphysical
free energy of the electronic system, which is
excited due to the broadening.

Using the scheme proposed by Methfessel and Paxton
in it's first order approximation leads
to energies, which are comparable to the energies obtained
by extrapolating the finite temperature energies to $T_{\rm el}=0$.
The calculated forces are the derivatives of these
energies and need no further correction.

% Put the label
% that you will use with \ref{} command in the braces 
% of the \label{} command.
%


\begin{references}
\bibitem{Chadi} D.J.~Chadi and M.L.~Cohen, 
Phys. Rev. B {\bf 8}, 5747 (1973).
\bibitem{Monkhorst} H.J.~Monkhorst and J.D.~Pack, Phys. Rev. B {\bf 12}, 5188 (1976).
\bibitem{FuHo} C.-L.~Fu and K.~.M.~Ho, Phys. Rev. B {\bf 28}, 5480 (1983).
\bibitem{Mermin} N.D.~Mermin, Phys. Rev. {\bf 137}, A 1441 (1969).
\bibitem{Davenport} M.~Weinert and J.W.~Davenport, Phys. Rev. B {\bf 45}, 13709 (1992).
\bibitem{ashcroft} N. W. Ashcroft and N. D. Mermin, 
{\sl Solid State Physics} (Saunders College, Philadelphia, 1976), pg.~47.
\bibitem{gillan} M. G. Gillan,
J. Phys. Condens. Matter {\bf 1}, 689 (1989).
\bibitem{jn} J. Neugebauer and M. Scheffler,
Phys. Rev. B {\bf 46}, 16067 (1992).
\bibitem{hamann} D. R. Hamann, Phys. Rev. B {\bf 40}, 2980 (1989).
\bibitem{gss} X. Gonze, R. Stumpf and M. Scheffler,
  Phys. Rev. B {\bf 44}, 8503 (1991).
\bibitem{chadi} D. J. Chadi and M. L. Cohen,
Phys. Rev. B {\bf 8}, 5747 (1973).
\bibitem{baroni1} S. Baroni, P. Giannozzi, and A. Testa,
Phys. Rev. Lett. {\bf 58}, 1861 (1987).
\bibitem{baroni2} S. Baroni, P. Giannozzi, and A. Testa,
Phys. Rev. Lett. {\bf 59}, 2662 (1987).
\bibitem{xg} X. Gonze and J.-P. Vigneron,
Phys. Rev. B {\bf 39}, 13120 (1989).
\bibitem{ws} A. Williams and J. Soler,
Bull. Am. Phys. Soc. {\bf 32}, 562 (1987).
\bibitem{payne} M. C. Payne, 
M. P. Teter, D. C. Allan, T. A. Arias,
and J. D. Johannopoulos, Rev. Mod. Phys. {\bf 64}, 1045 (1992).
\bibitem{meth} M. Methfessel and A. T. Paxton,
Phys. Rev. B {\bf 40}, 3616 (1989).
%\bibitem{tetra} O.~Jepsen and O.K.~Andersen, Solid State Comm.~9, 1763 (1971).
%\bibitem{bloechl} P. Bl\"ochl, O. Jepsen and O.K. Andersen,
%Phys. Rev. B {\bf 49}, 16223 (1994).
\end{references}
\end{document}